\documentclass[preprintnumbers, amsmath, amssymb, showpacs,twocolumn, superscriptaddress, 
nofootinbib]{revtex4-1}

\usepackage{graphicx}
\usepackage{epsfig}
\usepackage{dcolumn}
\usepackage{bm}
\usepackage{amsfonts}
\usepackage{subfigure}
\usepackage{color}

\newcommand{\be}{\begin{equation}}
\newcommand{\ee}{\end{equation}}

\newcommand{\bea}{\begin{eqnarray}}
\newcommand{\eea}{\end{eqnarray}}
\def\gsim{ \lower .75ex \hbox{$\sim$} \llap{\raise .27ex \hbox{$>$}} }
\def\lsim{ \lower .75ex \hbox{$\sim$} \llap{\raise .27ex \hbox{$<$}} }

\begin{document}

%---------------------------------------

\title{Implications of the possible 21-cm line excess at cosmic dawn on dynamics of interacting dark energy}

\author{Chunlong Li}
\email{chunlong@mail.ustc.edu.cn}
\affiliation{Department of Astronomy, School of Physical Sciences, University of Science and Technology of China, Hefei, Anhui 230026, China}
\affiliation{CAS Key Laboratory for Research in Galaxies and Cosmology, University of Science and Technology of China, Hefei, Anhui 230026, China}
\affiliation{School of Astronomy and Space Science, University of Science and Technology of China, Hefei, Anhui 230026, China}

\author{Xin Ren}
\email{rx76@mail.ustc.edu.cn}
\affiliation{Department of Astronomy, School of Physical Sciences, University of Science and Technology of China, Hefei, Anhui 230026, China}
\affiliation{CAS Key Laboratory for Research in Galaxies and Cosmology, University of Science and Technology of China, Hefei, Anhui 230026, China}
\affiliation{School of Astronomy and Space Science, University of Science and Technology of China, Hefei, Anhui 230026, China}

\author{Martiros Khurshudyan}
\email{khurshudyan@ustc.edu.cn}
\affiliation{Department of Astronomy, School of Physical Sciences, University of Science and Technology of China, Hefei, Anhui 230026, China}
\affiliation{CAS Key Laboratory for Research in Galaxies and Cosmology, University of Science and Technology of China, Hefei, Anhui 230026, China}
\affiliation{School of Astronomy and Space Science, University of Science and Technology of China, Hefei, Anhui 230026, China}
\affiliation{International Laboratory for Theoretical Cosmology, Tomsk State University of Control Systems and Radioelectronics (TUSUR), 634050 Tomsk, Russia}
\affiliation{Research Division,Tomsk State Pedagogical University, 634061 Tomsk, Russia}

\author{Yi-Fu Cai}
\email{yifucai@ustc.edu.cn}
\affiliation{Department of Astronomy, School of Physical Sciences, University of Science and Technology of China, Hefei, Anhui 230026, China}
\affiliation{CAS Key Laboratory for Research in Galaxies and Cosmology, University of Science and Technology of China, Hefei, Anhui 230026, China}
\affiliation{School of Astronomy and Space Science, University of Science and Technology of China, Hefei, Anhui 230026, China}

%---------------------------------------
\begin{abstract}
In this Letter we study implications of the possible excess of 21-cm line global signal at the epoch of cosmic dawn on the evolutions of a class of dynamically interacting dark energy (IDE) models. We firstly summarize two dynamical mechanisms in which different background evolutions can exert considerable effects on the 21-cm line global signal. One of them is the change in decoupling time of Compton scattering heating, the other stems from the direct change of optical depth due to the different expansion rate of the Universe. After that, we investigate the influence of linear IDE models on 21-cm line signals and find that under the current observational constraints, it is difficult to yield a sufficiently strong 21-cm line signal to be consistent with the results of Experiment to Detect the Global Epoch of reionization Signature (EDGES) since only the optical depth could be effectively changed in these models. Accordingly, this implies us to construct a background evolution which could fulfill the reasonable change of optical depth and Compton heating decoupling time at the same moment by introducing an early dark energy dominated stage into the evolution governed by the IDE models. The comparison with astronomical observations indicate that this scenario could only alleviate, but not complete eliminate, the tension between EDGES and other cosmological surveys. 
\end{abstract}

\pacs{98.80.-k, 98.80.Es, 95.36.+d, 95.36.+x}

\maketitle

\section{Introduction}

The underlying physics of the 21-cm line signal in the early Universe has become a hot topic since the Experiment to Detect the Global Epoch of Reionization Signature (EDGES) reported an excess of the 21-cm absorption line around the epoch of cosmic dawn. The strength of this signal is given by $T_{21}=-500^{+200}_{-500}mK$ at the redshift $z=17.2$ \cite{Bowman:2018yin}, which is $3.8\sigma$ below the strongest possible absorption under standard expectations $T_{21}=-0.209K$. 
It is known that the potential probe of these cosmological 21-cm lines from neutral hydrogen are significant to explore the epoch of reionization, which is almost invisible to other astronomical instruments (namely, see \cite{Pritchard:2011xb, Morales:2009gs, Furlanetto:2006jb} for comprehensive reviews). As a result, this observational anomaly has inspired extensive studies on the theoretical interpretations and phenomenological implications in the literatures \cite{Tashiro:2014tsa, Feng:2018rje, Barkana:2018cct, Mahdawi:2018euy, Mirocha:2018cih, Ewall-Wice:2018bzf, Hirano:2018alc, Venumadhav:2018uwn, Clark:2018ghm, Hektor:2018qqw, Safarzadeh:2018hhg, Munoz:2018jwq, Pospelov:2018kdh, Liu:2018uzy, Li:2018kzs, Lawson:2018qkc, Jia:2018csj, Hektor:2018lec, Yoshiura:2018zts, Jia:2018mkc, Houston:2018vbk, Chatterjee:2019jts, Boyarsky:2019fgp}. In addition, some discussions regarding the validity of the EDGES results have also been stimulated \cite{Bradley:2018eev, Nhan:2018cvm, Singh:2019gsv}.

Given that the brightness temperature of 21-cm line signal is defined by the difference between the background radiation temperature and the spin temperature of hydrogen atom, there are two straightforward methods to generate a possible strong signal. One is to enhance the background radiation through processes such as dark matter decay or annihilation \cite{Fraser:2018acy, Yang:2018gjd, DAmico:2018sxd, Mitridate:2018iag, Cheung:2018vww, Bhatt:2019qbq}, while the other is lowering the gas temperature by interactions between dark matter and baryons \cite{Barkana:2018lgd, Kovetz:2018zan, Slatyer:2018aqg, Munoz:2018pzp, Berlin:2018sjs, Barkana:2018qrx, Fialkov:2018xre}. However, most of these mechanisms would inevitably encounter some tensions when confronted with other astronomical observations. Accordingly, some novel scenarios were put forward which involve additional cooling or heating mechanisms induced by different species of the dark matter \cite{Li:2018kzs} and axions \cite{Houston:2018vrf, Auriol:2018ovo},  or the modification of the background evolution via Early Dark Energy EDE \cite{Hill:2018lfx} and Interacting Dark Energy (IDE) models \cite{Xiao:2018jyl, Costa:2018aoy, Wang:2018azy}.

In the present Letter we revisit the mechanisms on how different cosmological background evolutions could exert influence on the global 21-cm line signal in the early Universe. We point out that a specific background evolution would directly yield an impact on the optical depth of the hydrogen cloud and also the decoupling time of the Compton-heating process \cite{Hill:2018lfx}. Both could have considerable influence on the final strength of the 21-cm line signal at the epoch of cosmic dawn. Accordingly, in the present study we consider both aspects at the same time in order to investigate the possible implications for 21-cm line signal.

We start with the linear IDE models since they are regarded as the effective mechanisms of changing the evolution of Hubble parameter during the matter dominated era related to the 21-cm line signal at cosmic dawn \footnote{Cosmological models involving non-gravitational interactions between dark energy and dark matter were extensively studied in literatures \cite{Farrar:2003uw, Wang:2016lxa, Zimdahl:2001ar, Clemson:2011an, Bamba:2012cp, Bolotin:2013jpa, Li:2013bya, Li:2015vla, Zhang:2017ize, Kumar:2017dnp, Wetterich:1994bg,  Riess:2019cxk, Taubenberger:2019qna, Pan:2019gop, Cai:2019bdh, Yan:2019gbw, Lambiase:2018ows, Bhattacharyya:2018fwb, Kazantzidis:2018rnb}. For instance see \cite{Farrar:2003uw} for the early study, see \cite{Zimdahl:2001ar} for the alleviation of coincidence problem of the current cosmic acceleration, and see \cite{Wang:2016lxa} for a review. One particular motivation of this study is to realize an effective scenario for the equation-of-state parameter of dark energy across the cosmological constant boundary, which is dubbed as quintom cosmology \cite{Feng:2004ad, Cai:2006dm, Cai:2007gs, Xia:2007km, Cai:2007qw, Cai:2007zv, Cai:2008gk, Cai:2008ed, Cai:2012yf, Zhang:2005kj}. Additionally, we refer to \cite{Copeland:2006wr, Frieman:2008sn, Caldwell:2009ix, Cai:2009zp, DeFelice:2010aj, Cai:2015emx} for related reviews on various dynamical models driving the late-time cosmic acceleration.}. We examine whether these IDE models could be consistent with current observational constraints. Although there exists the severe tension between the limits of EDGES and other experiments, the analysis of what degree could the optical depth and Compton heating decoupling time be affected leads us to a more suitable form of the evolution for the Hubble parameter to be consistent with an anomalously strong 21-cm absorption feature. Then we fulfill this scenario by introducing a cosmological phase dominated by dynamical dark energy at early time and discuss the feasibility of this scenario under current observational constraints on the paradigm of IDE. Our analysis shows that, although this scenario can help to interpret an excess 21-cm line signal, the tension between EDGES and other astronomical constraints remains. We expect that this analysis could inspire the forthcoming consideration on the possible connection between an excess 21-cm line signal and the cosmic background evolution in a more reasonable way.

The structure of this Letter is as follows. In Section \ref{fundamental} we present a review of the global 21-cm line signal in the early Universe, pointing out that two mechanisms by which different background evolutions could affect 21-cm signal. In Section \ref{IDE}, we consider a class of linear IDE models and study if these models can be consistent with EDGES results using the current observational constraints from cosmic microwave background (CMB), baryon acoustic oscillations (BAO) and type Ia supernova (SNIa). We also apply the analyses of the optical depth and decoupling time of Compton scattering heating in the IDE models. In Section \ref{EDE}, we investigate the form of evolution for the Hubble parameter that could yield an excessive 21-cm signal by involving  the domination of dynamical dark energy at early stage. We then present our results along with further discussions in Section \ref{Conclusion}.

\section{21-cm line brightness temperature and the background evolution}
\label{fundamental}

The cosmological 21-cm line is caused by the hyperfine splitting of neutral hydrogen atoms, whose wavelength corresponds to the transition from the triplet state to the singlet state of the electron. We use the brightness temperature $T_{21}$ to describe the strength of the global sky-average signal, which is defined by the difference between the spin temperature $T_S$ of the hydrogen atom and the background radiation temperature $T_{\gamma}$ \cite{Ciardi:2003hg, Zaldarriaga:2003du}. Its form is expressed as follows,
\begin{align}
\label{T21}
 T_{21}=\frac{T_S-T_{\gamma}}{1+z}\big(1-e^{-\tau}\big) \approx\frac{T_S-T_{\gamma}}{1+z}\tau ~,
\end{align}
where $\tau$ is the optical depth of the diffuse inter-galactic medium
\begin{eqnarray}
\label{Tau}
 \tau=\frac{3}{32\pi}\frac{T_*}{T_S}n_{HI}\lambda_{21}^3\frac{A_{10}}{H(z)} ~.
\end{eqnarray}
In this formalism, $T_{*}$ corresponds to the energy of the 21-cm photon transition, $A_{10}$ is the downward spontaneous Einstein coefficient \cite{AliHaimoud:2010ab, AliHaimoud:2010dx}, $n_{HI}$ is the number density of neutral hydrogen and $\lambda_{21}$ is the wavelength of the 21-cm line. Due to the Wouthuysen-Field effect induced by the Ly$\alpha$ photons scattering within the gas at cosmic dawn, the spin temperature is approximately equal to the gas temperature, i.e. $T_{S}\simeq T_{b}$ \cite{Chen:2003gc, Wouthuysen:1952, Field:1958}.

In order to obtain the brightness temperature of the 21-cm signal, we need to know the evolution of the gas temperature $T_b$, which is determined by the Compton evolution equations \cite{AliHaimoud:2010ab, Seager:1999bc}:
\begin{align}
\label{gas}
 \frac{dT_b}{dz} (1+z)=2T_b+\frac{T_b-T_{\gamma}}{H t_C} ~,
\end{align}
where $T_{\gamma} = 2.725 (1+z) {\rm ~K}$ is the background radiation temperature and $t_C$ is the Compton-heating timescale, whose expression is given by
\begin{align}
\label{Compton time}
 t_C=\frac{3(1+f_{He}+x_e)m_e c}{8\sigma_T a_r T_{\gamma}^4 x_e} ~,
\end{align}	
where $\sigma_T$ is the Thomson scattering cross section, $a_r$ is the radiation constant, $m_e$ is the electron mass, $c$ is the speed of light, $f_{He}$ is the fractional abundance of helium by number and $x_e$ is the free electron fraction normalized to the hydrogen number density, i.e. $x_e=n_e/n_H$ and it evolves as \cite{AliHaimoud:2010dx}: 
\begin{align}
 \frac{dx_e}{dz} (1+z) = \frac{C_P}{H} \big[n_{H}A_B x_e^2-4(1-x_e) B_B e^{-\frac{3E_0}{4T_{\gamma}}} \big] ~,
\end{align}
where $E_0$ is the ground energy of hydrogen, $C_P$ is known as the Peebles $C$-factor, $A_B$ and $B_B$ are the effective recombination coefficients and the effective photoionization rate to and from the excited state, respectively. More detailed discussions on the underlying physics can be found in \cite{AliHaimoud:2010ab, AliHaimoud:2010dx}.
	
From the above description, it is obvious that the modification to the background evolution, i.e. a different evolution form of the Hubble parameter $H(z)$, shall alter the final brightness temperature of 21-cm line signal in two possible ways. One is that the Hubble parameter directly appears in the expression of the optical depth \eqref{Tau}, and hence a different value of the  Hubble parameter at a given redshift can change the brightness temperature of the 21-cm line at the corresponding redshift \cite{Pritchard:2011xb, Furlanetto:2006jb}. Specifically, if the value of the Hubble parameter at redshift $z=17.2$ were about 2/3 of that derived in the standard $\Lambda$CDM paradigm, then the signal of the 21-cm line can fall into the observed parameter space as claimed by EDGES.

The second effect comes from the second term on the r.h.s. of \eqref{gas}, which depicts the Compton scattering effects on the evolution of gas temperature. At $H(z)\simeq 1/t_C(z)$, the Compton scattering heating nearly decouples from gas temperature and the cooling law of the gas changes from $T_b\propto (1+z)$ to the pure adiabatic case $T_b\propto (1+z)^2$. Thus, an earlier time at which $H(z)\simeq 1/t_C(z)$ shall lead to a lower gas temperature at a given low redshift region. Then, according to the relation $T_{S} \simeq T_b$ at cosmic dawn, a stronger 21-cm line absorption signal would be obtained. Accordingly, we can estimate the decoupling time to be $z\simeq 161$ from the upper limit of EDGES's results, $T=-0.3K$, while in the standard $\Lambda$CDM cosmology this decoupling moment is estimated to be $z\simeq 120$.

We mention that, the neutral hydrogen number density $n_{HI}$ in Eq. \eqref{Tau} also seems to affect the 21-cm brightness temperature via a different background evolution. However, around the corresponding redshift ($z\simeq 17$) during the cosmic dawn, the recombination process had already finished, and thus this mechanism can hardly produce a signature of observable interest. 

\section{The IDE models and an excess of 21-cm line signal}
\label{IDE}

The thermal history of the Universe, being the most relevant aspect to the 21-cm line signal at cosmic dawn, starts from the recombination stage to some time near $z\sim 15$. In this period, the Universe was dominated by pressure-less matter, hence, we can approximate $H^2\approx(8\pi G/3)\rho_m$. Therefore, one mechanism to change the background evolution is to alter the amount of matter during this period. This is the key element in IDE models, which allows energy flow between dark matter and dark energy.

The possible interaction process between dark matter and dark energy can be parametrized through the continuity equations for their energy densities as follows,
\begin{align}
 &(1+z)H\frac{d\rho_{c}}{dz}-3H\rho_{c} = -Q ~, \nonumber\\
 &(1+z)H\frac{d\rho_{d}}{dz}-3H(1+\omega)\rho_d = Q ~,
\end{align}
where $\rho_{d}$ and $\rho_{c}$ represent the energy density of dark energy and cold dark matter, respectively. {, while $\omega$ and Q are the effective EoS parameter of dark energy and the non-gravitational interacting energy transfer respectively.  Different IDE models can be obtained by choosing different forms of $Q$. In the present study, we proceed our analysis by taking some phenomenological parameterized forms of $Q$. For simplicity, we consider the linear interaction forms for the IDE models as examples, which have been well studied in the literature \cite{He:2008tn}.
%\noteb{,which is a fairly common model and has been well studied} . 
The specific forms of the energy transfer $Q$ considered are,
\begin{align}
	{\rm Mode~ I-1}:\ &Q_{I-1}=3\lambda H\rho_d ~, \nonumber \\
	&-1<\omega<0 ~, \ \lambda<0 ~, \\
	{\rm Model~ I-2}:\ &Q_{I-2}=3\lambda H\rho_d ~, \nonumber \\
	&\omega<-1 ~, \ 0<\lambda<-2\omega\Omega_c ~, \\
	{\rm Model~ II}:\ & Q_{II}=3\lambda H\rho_c ~, \nonumber \\ 
	&\omega<-1~, \ 0<\lambda<-\omega/4 ~, \\
	{\rm Model~ III}:\ & Q_{III}=3\lambda H(\rho_d+\rho_c) ~, \nonumber \\
	&\omega<-1 ~, \ 0<\lambda<-\omega/4 ~,
\end{align}
where $\Omega_c$ is the density parameter of cold dark matter. We mention that, the choice of the parameter space of the IDE models is expected to avoid the instability at perturbation level. 
%When choosing some range of the parameters in IDE model, it may lead the instabilities in perturbation level. 
Based on the stability analyses of perturbations \cite{He:2008si, Deffayet:2010qz, Yang:2018euj}, there exist desirable parameter space for EoS parameter of dark energy $\omega$ and the interaction parameter $\lambda$ to ensure the stability of the models. 
This issue can also be addressed by the so-called parametrized post-Friedmann approach. This approach has been used to calculate the perturbation equations of IDE models, where large-scale instability can be avoided in general IDE models and a wide range of parameter space is available \cite{Zhang:2017ize, Dai:2019vif}. Note that, ${\rm Model~ I-1}$ and ${\rm Model~ I-2}$ have the same form of interaction term but different allowed parameter spaces.

\subsection{The EDGES's results v.s. other cosmological constraints}
\label{constraints}

The aforementioned four models have been comprehensively studied and well constrained in the work of \cite{Costa:2016tpb} by using the data from Planck 2015, baryon acoustic oscillations (BAO) and Type Ia supernovae (SNIa). The related analyses can also be found in the literature \cite{He:2010im, Santos:2017bqm, Yang:2018pej, Li:2018ydj, LeDelliou:2018vua}. Table \ref{models} provides the main constraints on the model parameters from \cite{Costa:2016tpb}, which closely relates to our discussion. In the following study, we shall compare these constraints with the results derived from EDGES to see whether if a reasonable IDE model would be consistent with an excess in the 21-cm global signal.

\begin{table*}
\centering
\begin{tabular}{c|cccc}
\hline
Model & $\omega$ & $\lambda$ & $\Omega_c h^2$ & $H_0$ \\
\hline
I-1: $Q_{I-1}=3\lambda H \rho_d$ & $-0.9191^{+0.0222}_{-0.0839}$ & $-0.1107^{+0.085}_{-0.0506}$ & $0.0792^{+0.0348}_{-0.0166}$ & $68.18^{+1.43}_{-1.44}$ \\
I-2: $Q_{I-2}=3\lambda H \rho_d$ & $-1.088^{+0.0651}_{-0.0448}$ & $0.05219^{+0.0349}_{-0.0355}$ & $0.1351^{+0.0111}_{-0.00861}$ & $68.35^{+1.47}_{-1.46}$ \\
II: $Q_{II}=3\lambda H\rho_c$ & $-1.104^{+0.0467}_{-0.0292}$ & $0.0007127^{+0.000256}_{-0.000633}$ & $0.1216^{+0.00119}_{-0.00119}$ & $68.91^{+0.875}_{-0.997}$ \\
III: $Q_{III}=3\lambda H(\rho_d+\rho_c)$ & $-1.105^{+0.0468}_{-0.0288}$ & $0.000735^{+0.000254}_{-0.000679}$ & $0.1218^{+0.00125}_{-0.00133}$ & $68.88^{+0.854}_{-0.97}$ \\
\hline
\end{tabular}
\caption{The latest cosmological constraints on model parameters of the IDE paradigm at 68\% C.L. as derived from \cite{Costa:2016tpb}.}
\label{models}
\end{table*}

 We fix today's Hubble parameter since its uncertainty is too small to make a difference on the following results, and we also pick up the three most relevant parameters $\omega$, $\lambda$ and $\Omega_ch^2$. Then we calculate the corresponding 21-cm line brightness temperature at the redshift $z=17.2$. The results are displayed in Figure \ref{experiments}. In these figures, we plot the boundary values of the constraints on $\omega$, $\lambda$ and $\Omega_ch^2$ from Table \ref{models}. The parameter spaces that lie bottom-right relative to the lines could give rise to a 21-cm brightness temperature signal that is stronger than the upper limit of the EDGES result $T_{21}=-0.3K$, and could therefore be supported by this experiment.

\begin{figure}
\includegraphics[width=.4\textwidth]{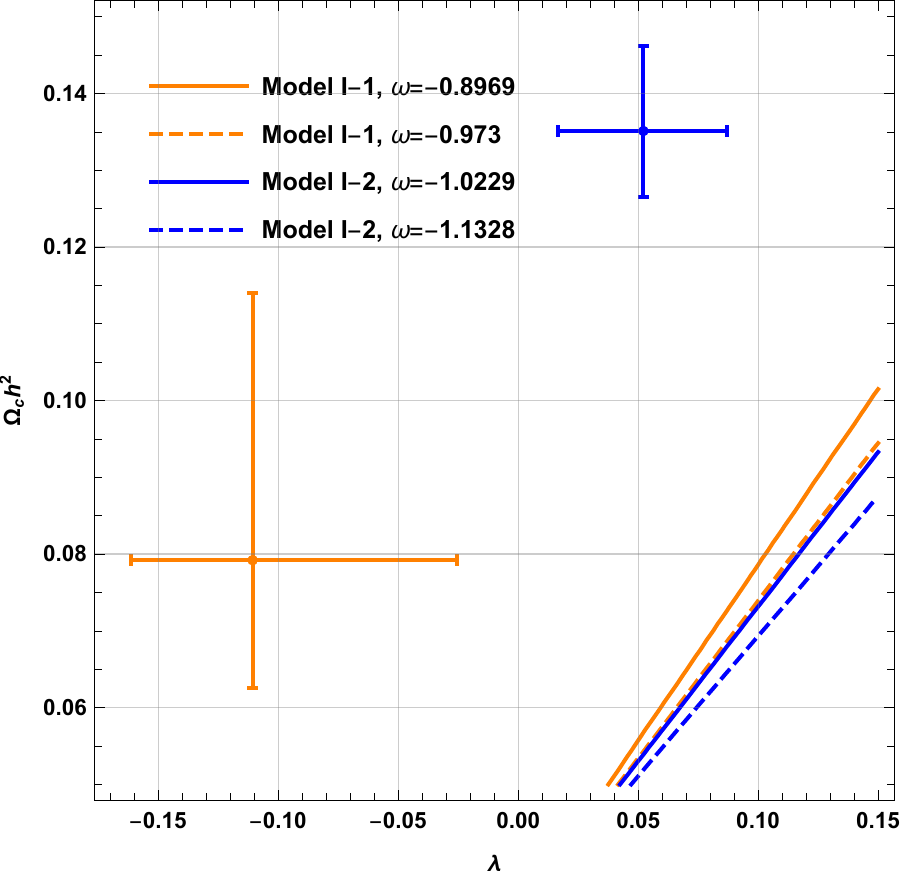}
\includegraphics[width=.4\textwidth]{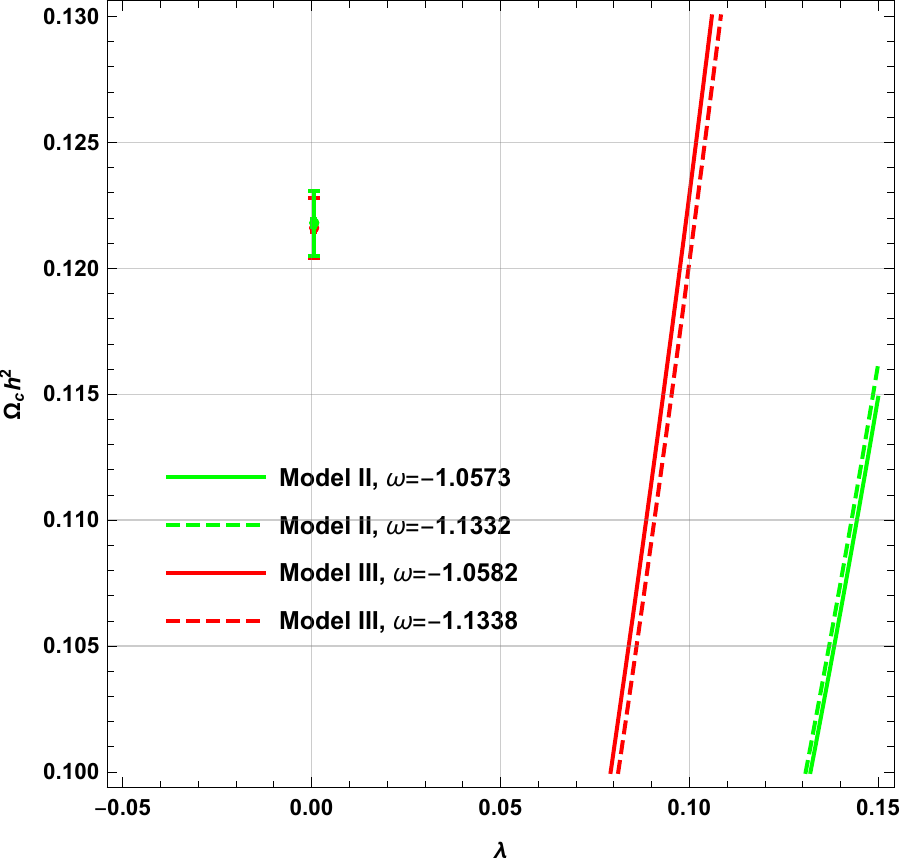}
\caption{Comparison between EDGES and other experiments for different IDE models with different EoS parameters $\omega$. The values of $\lambda$ and $\Omega_ch^2$ picked up by the lines yield the upper limit of EDGES's results $T_{21}=-0.3K$. The parameter spaces that lie right and below the lines can lead to a stronger 21-cm brightness temperature signal. The error bars are derived from the constraints on $\lambda$ and $\Omega_ch^2$ and their colours correspond to different IDE models as has been explained in the plot.}
\label{experiments}
\end{figure}

The constraints for parameter $\Omega_ch^2$ and $\lambda$ from Table \ref{models} are labeled by error bars of the same colors as their corresponding models. Note that in the second panel of Figure \ref{experiments}, the constraints are very tight for Model II and Model III. As pointed out in \cite{He:2010im}, these two models would significantly alter the CMB power spectrum at low $\ell$ and hence are tightly constrained. As a result, we can see that a tension exists between the limit of the EDGES and other experiments for the IDE model with a linear interaction term.

\subsection{Mechanisms of affecting the global 21-cm lines}

Although the IDE models with a linear interaction term seem to be inconsistent with an excess 21-cm line signal reported by EDGES, it is still interesting to study the implications of an abnormal 21-cm signal on the evolution of the cosmological background. In the following section we will explore in detail the mechanisms in which the IDE models can affect the global 21-cm signal.
As we have mentioned, the background evolution can alter the signal of the 21-cm brightness temperature in two possible ways. One is the direct change of the optical depth, and the other is changing the decoupling time of Compton heating.

In Figure \ref{Compton}, we show $H(z)$ and the Compton-heating rate $1/t_C(z)$ for the $\Lambda$CDM model and different IDE models with different interacting parameters. Around $z\sim 17$, different cases have different values of Hubble parameter. So according to Eq. (\ref{Tau}), change of optical depth would result in some observable effects and a smaller value of the Hubble parameter tends to give rise to a stronger brightness temperature. As for the effects from the change in Compton scattering decoupling time, we notice that} the time at which $H(z)\simeq 1/t_C(z)$, i.e. the intersection of the solid and dashed lines in this plot marks, the decoupling time of Compton-heating for each case. As mentioned above, an earlier presence of the intersection could help producing a stronger brightness signal. On the other hand, although the IDE could change the evolution of $H(z)$ and $1/t_{C}$, the total effect only changes the decoupling time at that $H(z)\simeq 1/t_C(z)$  a little. So we expect the main contribution to the change of the 21-cm brightness temperature would be from the change of the optical depth instead of the Compton heating decoupling time.

\begin{figure}
\includegraphics[width=.4\textwidth]{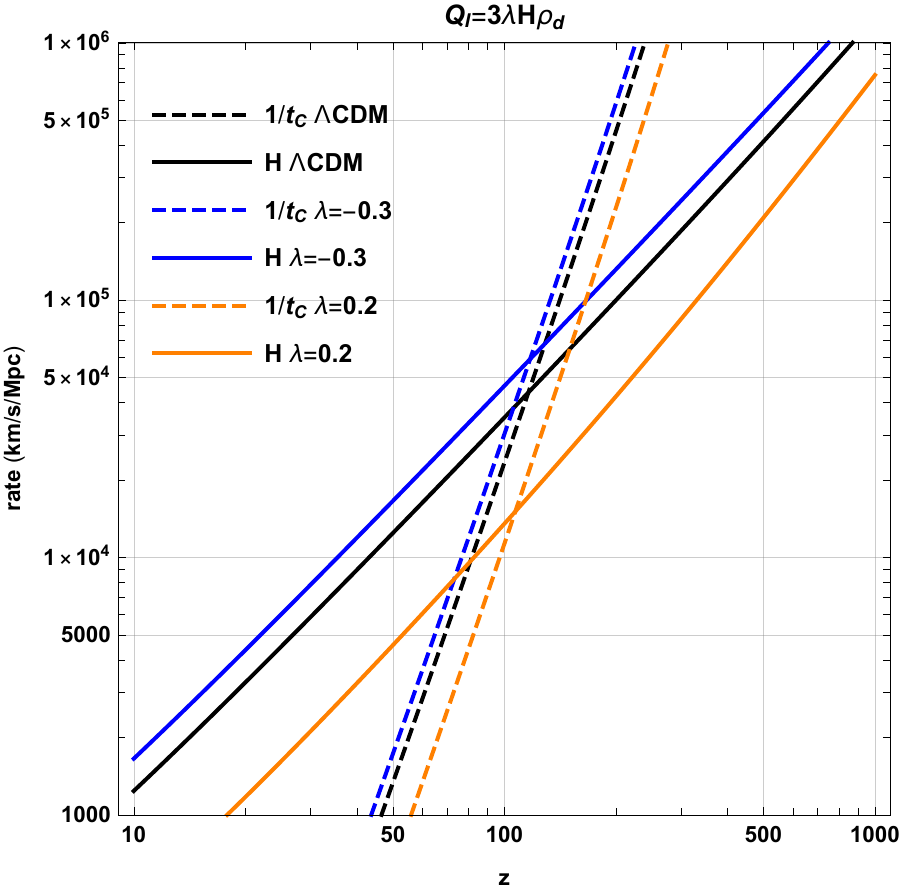}
\includegraphics[width=.4\textwidth]{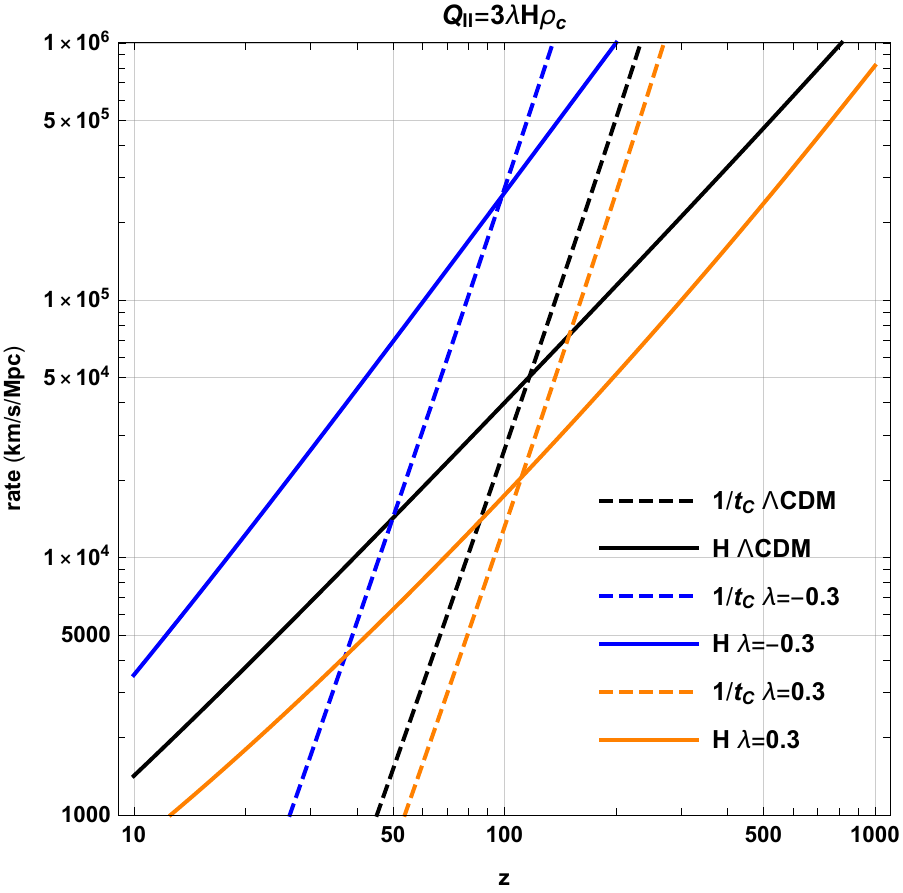}
\includegraphics[width=.4\textwidth]{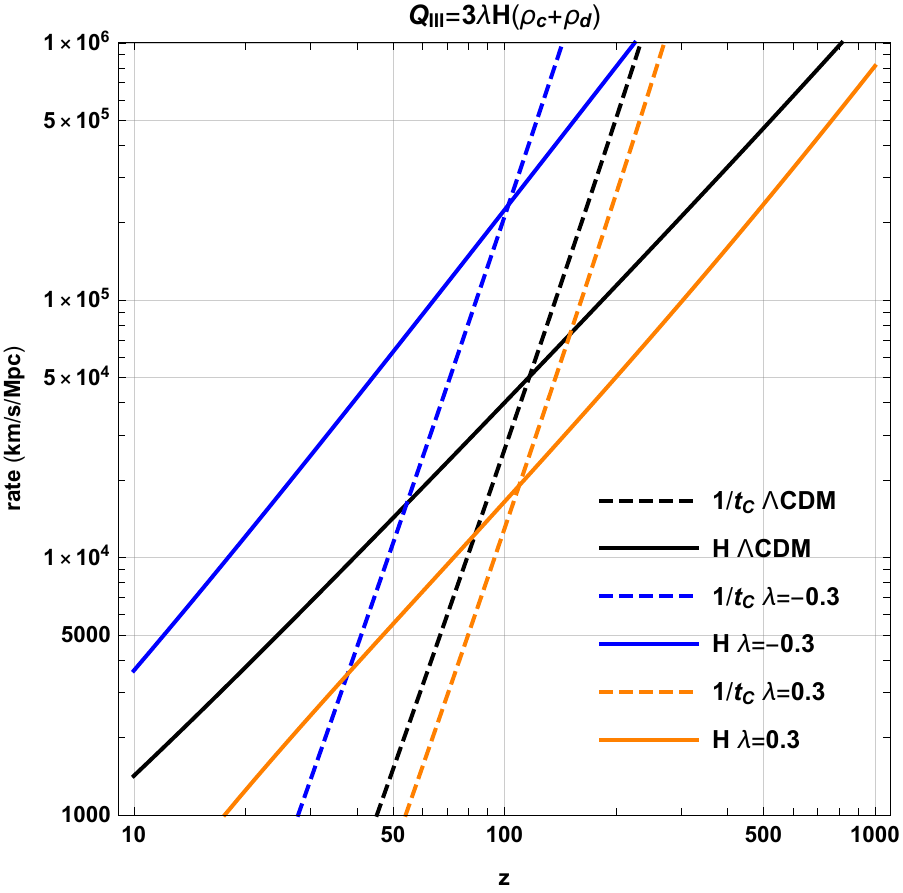}
\caption{The Hubble parameter and Compton-heating rate for different interacting strengths and different IDE models. The decoupling of the gas temperature from the radiation temperature occurs when $H(z)\approx 1/t_C(z)$ for a given model and a given $\lambda$, i.e. the intersection of the lines with the same color.}
\label{Compton}
\end{figure}

\begin{figure}
\includegraphics[width=.4\textwidth]{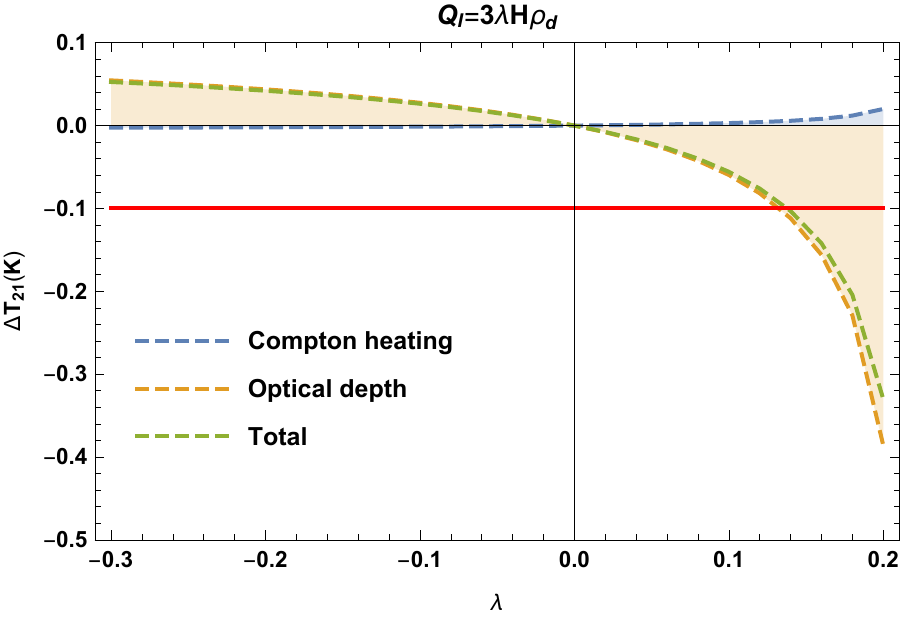}
\includegraphics[width=.4\textwidth]{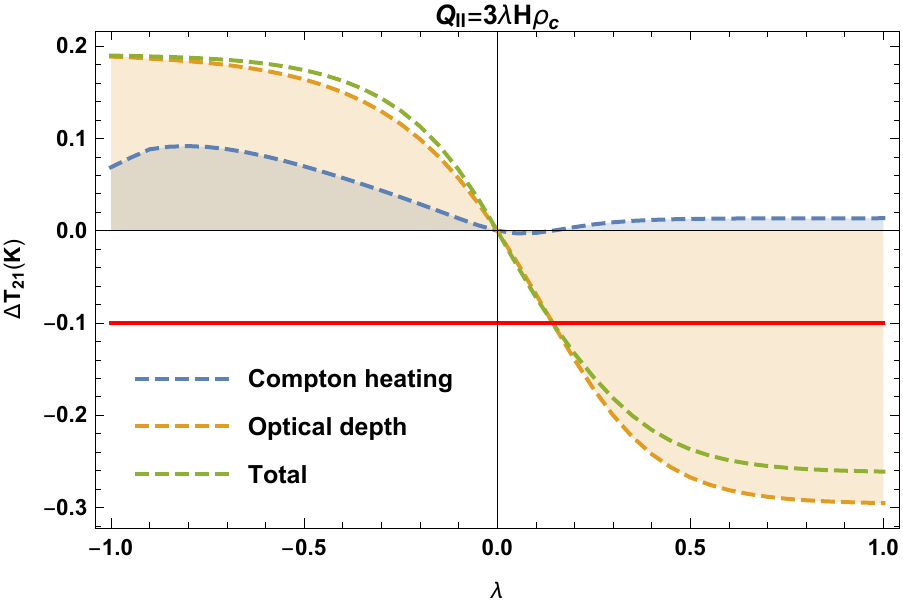}
\includegraphics[width=.4\textwidth]{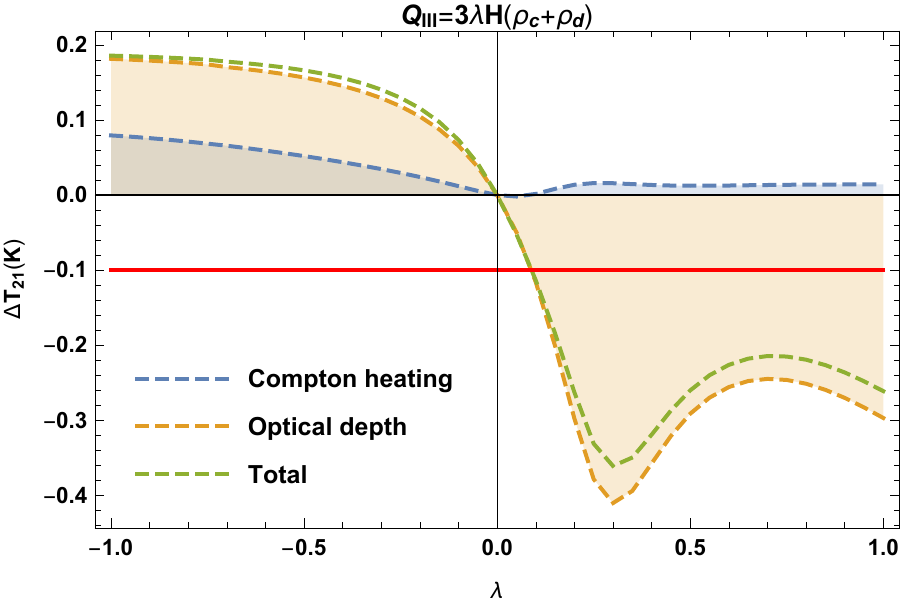}
\caption{The change of the 21-cm line signal $\Delta T_{21}$ from different factors (Compton-heating, cosmic expansion and taking both of them into consideration.) for different models. And we use the red line to label the upper limit of EDGES's results $\Delta T_{21}=-0.1K$. The areas surrounded by dashed lines and $\Delta T_{21}=0$ measure the degree of influence of different factors. Note that the parameter range of $\lambda$ is slightly different among different models, since we need to take the singularity of the models into our consideration.}
\label{delta T21}
\end{figure}

To better demonstrate the effect of two mechanisms clearer, we define the change of 21-cm brightness temperature as $\Delta T_{21}=T_{21}^*-T_{21}^0$, where  $T_{21}^0\simeq -0.2K$ is the output value of 21-cm brightness temperature at $z=17.2$ for the standard $\Lambda$CDM model, $T_{21}^*$ is the corresponding value for different factors (optical depth and Compton heating) and different models. In Figure \ref{delta T21}, we plot the values of $\Delta T_{21}$ as a function of the interacting parameter $\lambda$ after considering different factors for different models. We also plot the upper limit of the EDGES result with the red line, i.e. $\Delta T_{21}=-0.1K$, and therefore, the parameter space that makes the $\Delta T_{21}$ below the red line is consistent with the EDGES results at 99\% C.L. .

From Figure \ref{delta T21}, we notice that a positive interacting parameter could lead to a relatively stronger 21-cm brightness temperature signal than that in $\Lambda$CDM model, and thus, would help alleviating the tension between standard cosmology and the observations of EDGES. Moreover, by comparing with the decoupling time of Compton heating, the change of optical depth has a larger influence on the signal. it is consistent with our previous analysis that the decoupling time of Compton heating could hardly be changed and a smaller Hubble parameter tends to result in a stronger 21-cm line signal from Eq. \eqref{Tau}.

\section{An early dark energy dominated stage}
\label{EDE}

According to the previous analysis, we can learn that the IDE models can only yield significant effects by changing the optical depth. However, given that there are two factors that could influence the 21-cm line signal, the best choice might be changing the decoupling time of Compton heating and the optical depth of hydrogen cloud at the same time.

\begin{figure}
\includegraphics[width=.4\textwidth]{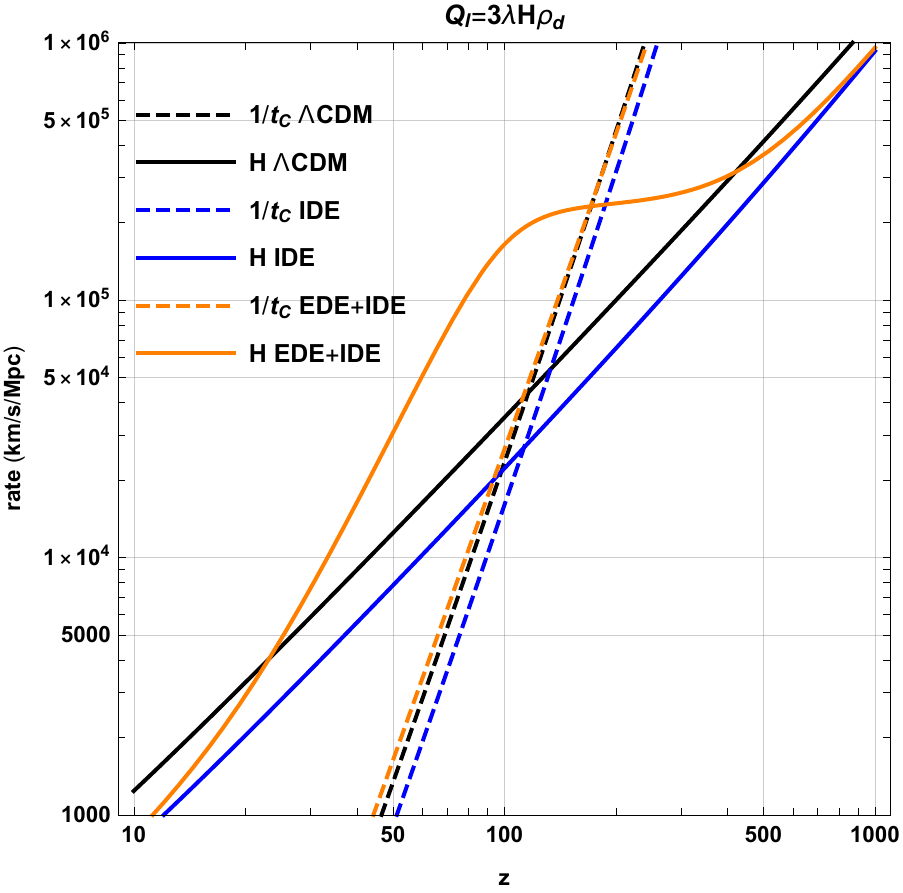}
\caption{The Hubble parameter and Compton-heating rate for $\Lambda$CDM, the IDE model $\lambda=0.15$ and the interacting plus early dark energy model $\lambda=0.15$, $\Omega_{ee}=0.5\times 10^{-5}$, $z_c=100$.}
\label{IEDE}
\end{figure}

In order to significantly change the decoupling time of Compton heating, i.e the intersection of $H(z)$ and $1/t_C(z)$, we construct a smooth evolution stage of $H(z)$ at the redshift $z\sim 100$. As \cite{Hill:2018lfx} points out, this scenario could be fulfilled with an early dark energy model. It can be expressed as \cite{Karwal:2016vyq, Poulin:2018cxd}:
\begin{align}
 \frac{\rho_{ee}(a)}{\rho_{crit}} =	\frac{\Omega_{ee}(1+a_c^6)}{a^6+a^6_c} ~,~ 
 p_{ee}(a) =\rho_{ee}\frac{a^6-a_c^6}{a^6+a_c^6} ~,
\end{align}
where $\rho_{crit}$ is the critical density at $z=0$, while $\Omega_{ee}$ and $a_c$ are the model parameters. For $z\gg z_c$, this new composite behaves as a cosmological constant $\omega=-1$ while for $z\ll z_c$, $\omega=1$ and the energy density approaches to $\Omega_{ee}\rho_{crit}$. We add the new early dark energy component to the IDE paradigm, we yield $H^2\approx (8\pi G/3)(\rho_m+\rho_{ee})$. Figure \ref{IEDE} displays the Hubble parameter and Compton-heating rate for the IDE plus early dark energy model, where we can see the evolution of the cosmological background can both lower the Hubble parameter at $z\sim 17$ and significantly push the decoupling time of Compton heating to an earlier time. Here we would like to comment that the form depicted by the ``EDE+IDE" in Figure \ref{IEDE} might offer a possible solution to alleviate the tension between the EDGES results and other cosmological observations.

%{\bf However, when we take the current observational constraints into our considerations, the situation might not be so optimistic.}
%
When we introduce the early dark energy dominated stage, the expansion history of the IDE paradigm changes. Given that we need the model parameter $z_c>10$ to realize the aim of changing the Compton-heating decoupling time only at relatively large redshift, a significant constraint is the precise measurement of the acoustic scale by CMB experiments, which is given by
\begin{align}
 \theta_*\equiv \frac{r_s(\eta_*)}{\eta_0} ~,
\end{align}
where $\eta_0$ is the comoving angular diameter distance to the surface of last scattering, $r_s(\eta_*)$ is the comoving sound horizon at the recombination stage. Focusing on the contribution of an early exponential expansion background after the recombination, the comoving sound horizon $r_s(\eta_*)$ would not be affected. Afterwards, by taking the current tight constraints on $\theta_*$ into consideration \cite{Costa:2016tpb, Aghanim:2018eyx}, our strategy is to keep $\eta_0$ constant, which turns out to yield an integral constraint on $H^{-1}(z)$. For the other parameters, we choose the suitable values to give rise to the strongest signal consistent with the experimental constraints given by Section \ref{constraints}.

If we add the new early dark energy component to the cosmological paradigm, the only way to keep $\eta_0$ constant is to alter the present Hubble parameter $H_0$ since the other parameters have been completely fixed. Furthermore, the uncertain range of $H_0$ should also be within the constraints provided in Table \ref{models}, which would then give rise to a rough constraint on $\Omega_{ee}$ and $a_c$.

\begin{figure*}
\includegraphics[width=.45\textwidth]{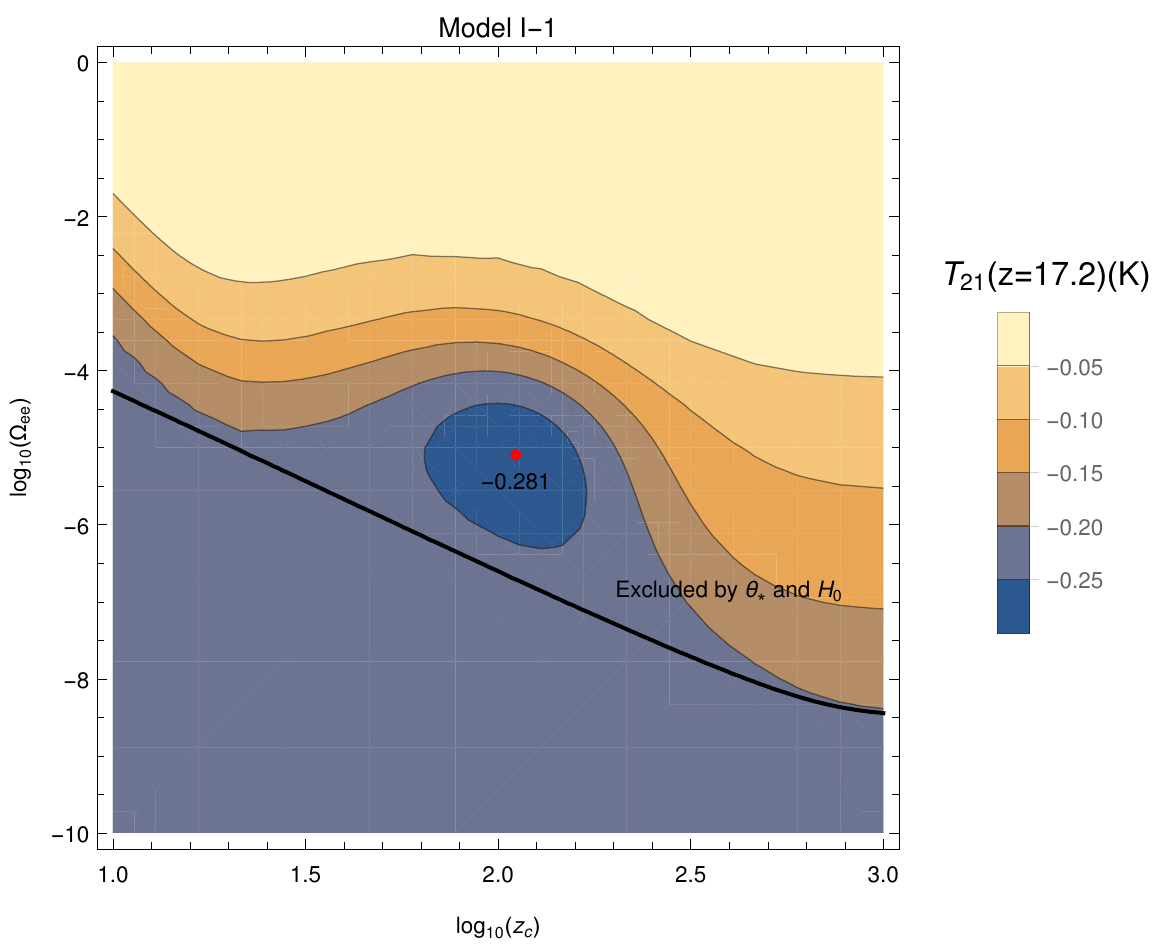}
\includegraphics[width=.45\textwidth]{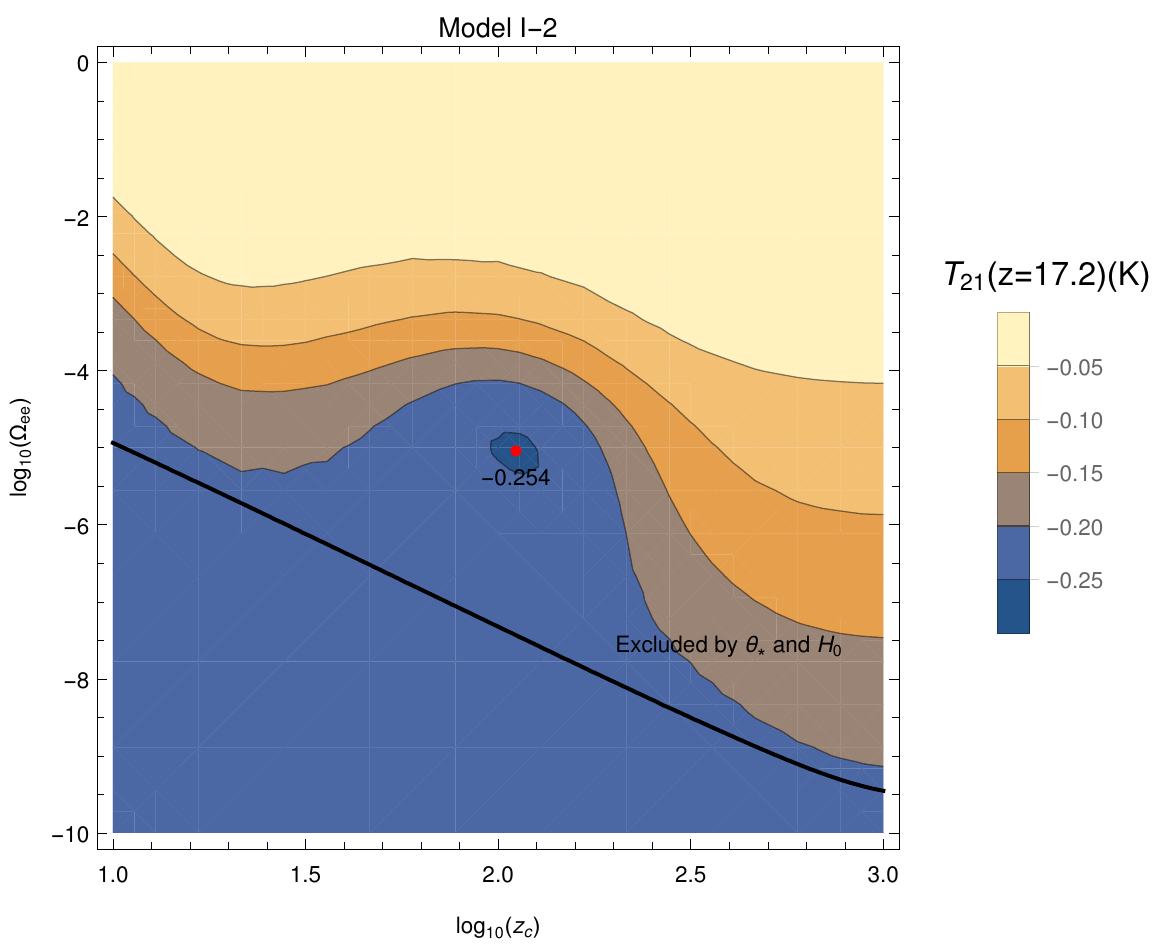}
\includegraphics[width=.45\textwidth]{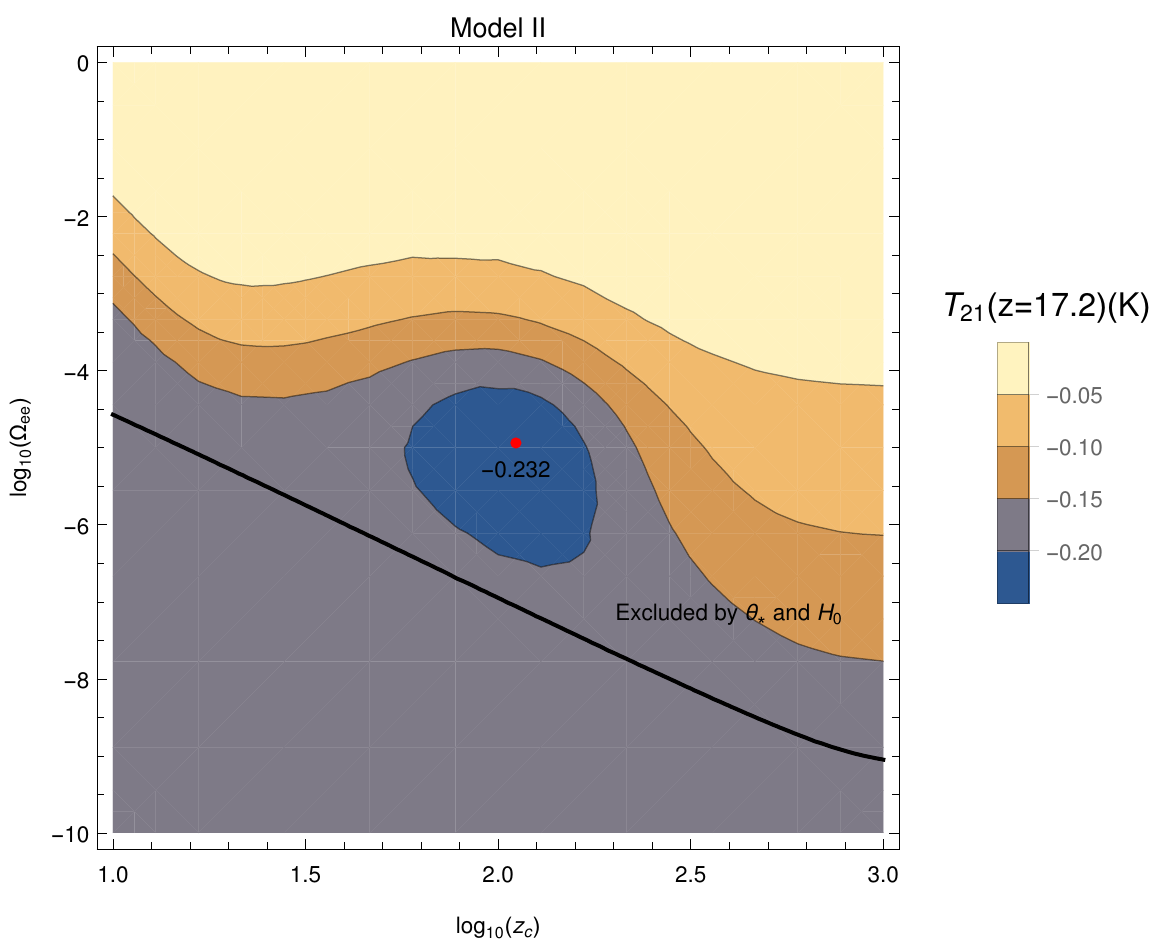}
\includegraphics[width=.45\textwidth]{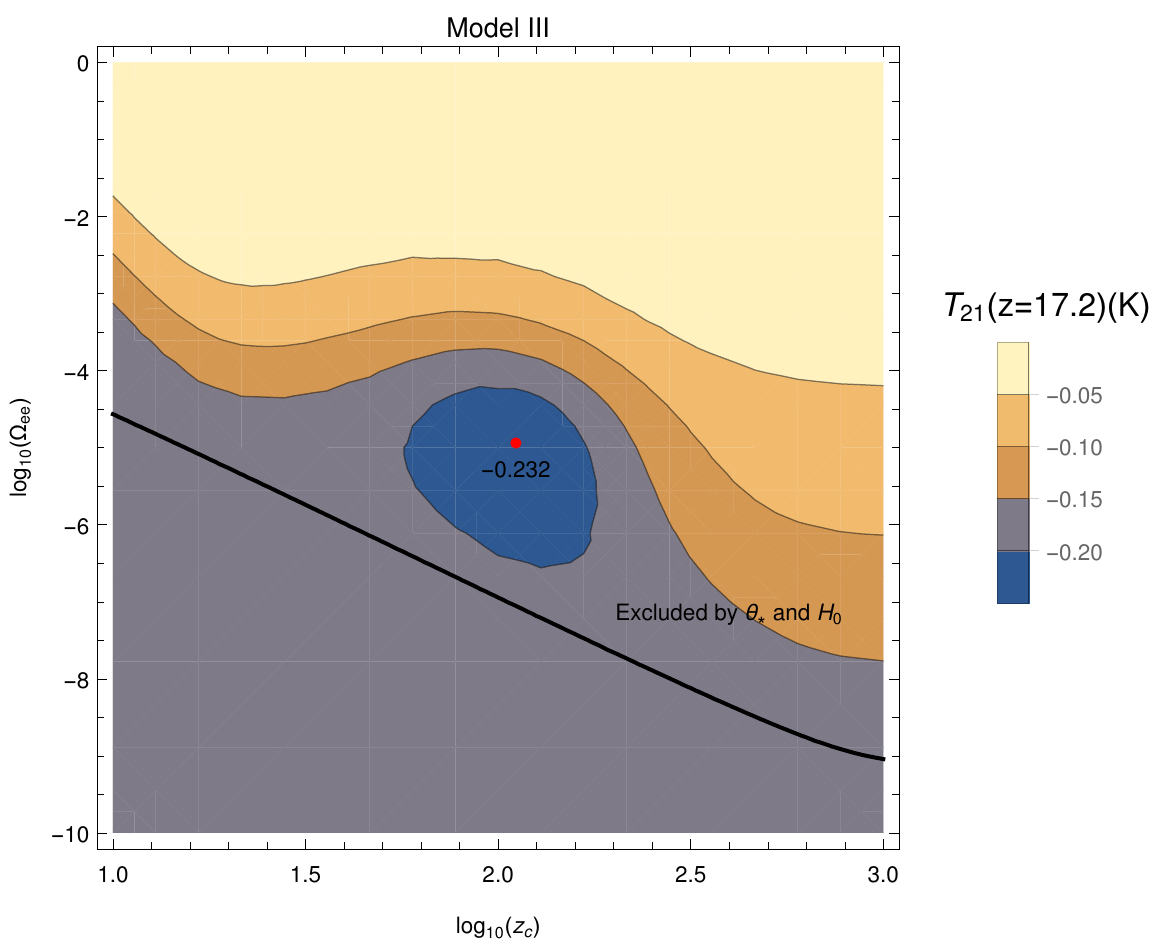}
\caption{The 21-cm line brightness temperature given by different early dark energy parameters $\Omega_{ee}$ and $z_c$ for different IDE models. The parameter spaces above the black line are excluded by the measurements of $\theta_*$ and $H_0$.}
\label{EDE results}
\end{figure*}

The results are shown in Figure \ref{EDE results}. Note that, the contours here tend to form  a circle, which is slightly different from the work of \cite{Hill:2018lfx} where the variation of the optical depth was ignored. If we have a larger $\Omega_{ee}$, the early dark energy effect will extend to lower redshift, which enhances 21-cm signal according to Eq. (\ref{Tau}). From these results, we find that by adding an early dark energy dominated stage to the IDE paradigm within the current observational constraints, the upper limit of EDGES's results $T_{21}=-0.3K$ at 99\% C.L. cannot be easily reached. Moreover, the parameter spaces providing the strongest brightness temperature signal might already be excluded by current constraints. The evolution form of Hubble parameter shown in Figure \ref{IEDE} by combining the IDE model and early dark energy model is still inconsistent with the large signal reported by EDGES.

Another crucial constraint arise from the observation of the CMB power spectra. If there is an early dark energy dominated stage playing a role during the period after the CMB having been formed, the evolution of the gravitational potential would be significantly modified due to a different growth function from that of a matter dominated stage. This would bring a considerable contribution to the integrated Sachs-Wolfe (ISW) effects, and therefore, the ``EDE+IDE" model would face a severe constraint from the CMB power spectra especially at the large scales (small $l$'s region).

Specifically, let us consider $z_c=100$ and $\Omega_{ee}=10^{-5}$ as the strongest 21-cm line signal is given by the parameter space around this point. In \cite{Karwal:2016vyq}, the authors numerically calculated the partial derivatives of the TT spectrum with respect to $\Omega_{ee}$ under the model of ``CDM+EDE" (FIG. 4 of \cite{Karwal:2016vyq}). Although these two models are different from each other, we still can make a magnitude estimation since their differences are sub-dominant given that these models can be well constrained. For the value of TT spectrum at $l=100$, the change of $D_{l}^{TT}$ is at least around $400(\mu K)^2$, which is much beyond the measurement error of the Planck satellite at this point. The same situation also occurs in the other low $l$ region and even more serious. So one may expect the parameter space that could give rise to the strongest 21-cm line signal are also excluded by the CMB observations.

\section{Conclusions}	
\label{Conclusion}

In this Letter, we studied the implications of the possible excess of the 21-cm global signal around cosmic dawn on the cosmological expansion of the early Universe. Especially, we point out the most suitable evolution of the Hubble parameter that is able to enhance the 21-cm signal by reviewing two potential mechanisms in which a different background evolution could yield impacts on the brightness temperature of the 21-cm line. One of them is the change of optical depth in the diffuse inter-galactic medium, while the other is the change in the Compton scattering decoupling time. We consider IDE models with the interacting term $Q=3\lambda H\rho_d$, $3\lambda H\rho_c$ and $3\lambda H(\rho_d+\rho_c)$ to demonstrate our analyses. By comparing with the current experimental data and analyzing two different mechanisms, we find that the linear IDE models only have observable effects through the change in optical depth. As a result, we derive the required evolution form for the Hubble parameter in order to realize the excessive 21-cm line signal, which includes a smaller value at $z\sim 17$ than that of the standard cosmology and an early smooth evolution stage around $z\sim 100$. This form of evolution could be obtained by adding an early dark energy dominated stage to the IDE paradigm. Finally, we consider the current experimental constraints from CMB, BAO and SNIa on the parameter space. Although the results show that this kind of model is still not efficient to yield a strong enough signal reported by EDGES, our study clearly reveals the possible connections between an excess 21-cm line signal at cosmic dawn and the underlying cosmic background evolution, which should inspire the community to find more novel ways to understand an excess 21-cm line signal.

With the large uncertainty of the EDGES measurement in mind, it remains difficult to make decisive conclusion on the cosmological models that were put forward to explain the excess of the 21-cm lines due to the severe tension with other astronomical experiments. In addition, we would also like to mention that there exist some debates about the background noise uncertainties of the EDGES observations \cite{Bradley:2018eev, Nhan:2018cvm, Singh:2019gsv}, which implies that more accurate signals of 21-cm line are expected. If 21-cm line signals from dark age can be measured precisely, this will be a brand new observational window for us to explore physics of the early Universe. In order to shed light on the mysterious period of the cosmic dawn, we hope for more precise astronomical surveys on the scan of 21-cm lines, such as the square kilometre array (SKA) \cite{Farnes:2018byc, Li:2019bsg} or other related projects, in the near future.

\begin{acknowledgments}
We thank Emilio Elizalde, Zhiqi Huang, Antonino Marciano, Hong Tsun Wong, Supriya Pan, Emmanuel Saridakis, Bin Wang, Dong-Gang Wang, Yuting Wang, Weiqiang Yang and Xin Zhang for stimulating discussions. 
YFC acknowledges to Professor Bin Wang for kind hospitality when attending the {\it HouYi} cosmology workshop at Yangzhou University. 
This work is supported in part by the National Youth Thousand Talents Program of China, by the NSFC (Nos. 11961131007, 11722327, 11653002, 11421303), by CAST Young Elite Scientists Sponsorship Program (2016QNRC001), and by the Fundamental Research Funds for the Central Universities. 
MK is supported in part by a CAS Presidents International Fellowship Initiative Grant (No. 2018PM0054) and the NSFC (No. 11847226). 
All numerical analyses are operated on the computer clusters ``{\it Linda \& Judy}'' in the particle cosmology group at USTC. 
\end{acknowledgments}

%-----------------------------------

\end{document}